\documentclass[conference]{IEEEtran}
\pdfoutput=1
\IEEEoverridecommandlockouts
\usepackage{cite}
\usepackage{amsmath,amssymb,amsfonts}
\usepackage{graphicx}
\usepackage{textcomp}
\usepackage{xcolor}
\usepackage{subfig}

\begin{document}

\title{Deep Learning for Multi-View Ultrasonic Image Fusion\\
\thanks{This work was supported by Applus+ RTD, CWI and the Dutch Research Council (NWO 613.009.106). }}
\author{
    \IEEEauthorblockN{Georgios Pilikos\IEEEauthorrefmark{1}, Lars Horchens\IEEEauthorrefmark{2}, Tristan van Leeuwen\IEEEauthorrefmark{1} and Felix Lucka\IEEEauthorrefmark{1}}
    \IEEEauthorblockA{\IEEEauthorrefmark{1}Computational Imaging, Centrum Wiskunde \& Informatica, Amsterdam, NL
    }
    \IEEEauthorblockA{\IEEEauthorrefmark{2}Applus+ E\&I Technology Centre, Rotterdam, NL}
}

\maketitle

\begin{abstract}
Ultrasonic imaging is being used to obtain information about the acoustic properties of a medium by emitting waves into it and recording their interaction using ultrasonic transducer arrays. The Delay-And-Sum (DAS) algorithm forms images using the main path on which reflected signals travel back to the transducers. In some applications, different insonification paths can be considered, for instance by placing the transducers at different locations or if strong reflectors inside the medium are known a-priori. These different modes give rise to multiple DAS images reflecting different geometric information about the scatterers and the challenge is to either fuse them into one image or to directly extract higher-level information regarding the materials of the medium, e.g., a segmentation map. Traditional image fusion techniques typically use ad-hoc combinations of pre-defined image transforms, pooling operations and thresholding. In this work, we propose a deep neural network (DNN) architecture that directly maps all available data to a segmentation map while explicitly incorporating the DAS image formation for the different insonification paths as network layers. This enables information flow between data pre-processing and image post-processing DNNs, trained end-to-end. We compare our proposed method to a traditional image fusion technique using simulated data experiments, mimicking a non-destructive testing application with four image modes, i.e., two transducer locations and two internal reflection boundaries. Using our approach, it is possible to obtain much more accurate segmentation of defects.
\end{abstract}


\section{Introduction}
Ultrasonic imaging generates maps of the acoustic properties of a medium by measuring the response to ultrasonic waves emitted into it. The Delay-And-Sum (DAS) image formation algorithm computes images by considering the main path of the reflected data to the transducers. To obtain complementary information, additional insonification paths can be utilized, involving the transmission and reception of waves using transducer arrays positioned at different locations. Boundary reflections can also be incorporated in the imaging method as well as direct insonification \cite{1}. 

An image per insonification path can be formed using the DAS algorithm to create multiple image modes, where each provides different geometric information about the scatterers. In some applications, there can be many images corresponding to the same region-of-interest. Fusing these image modes is not always trivial and often further interpretation and human expertise is required to extract the necessary information. Traditional fusion algorithms process images using ad-hoc combinations of pre-defined image transforms, pooling operations and thresholding to obtain higher-level information regarding materials of interest, e.g. a segmentation map. 

Recently, deep learning methods have been introduced to tackle this problem. They use data as input and output images by learning optimal parameters for beamforming \cite{2} \cite{3} \cite{4}. This was extended to receive raw data as input and learn how to create a beamformed image and a segmentation map simultaneously \cite{5}. Purely data-driven methods are successful but it was shown that the final results could be improved when combined with traditional physics-based image formation algorithms. This was demonstrated by incorporating the DAS image formation algorithm within deep learning training for both segmentation and imaging \cite{6} \cite{7}.

\begin{figure*}
\centering
\includegraphics[scale=0.13]{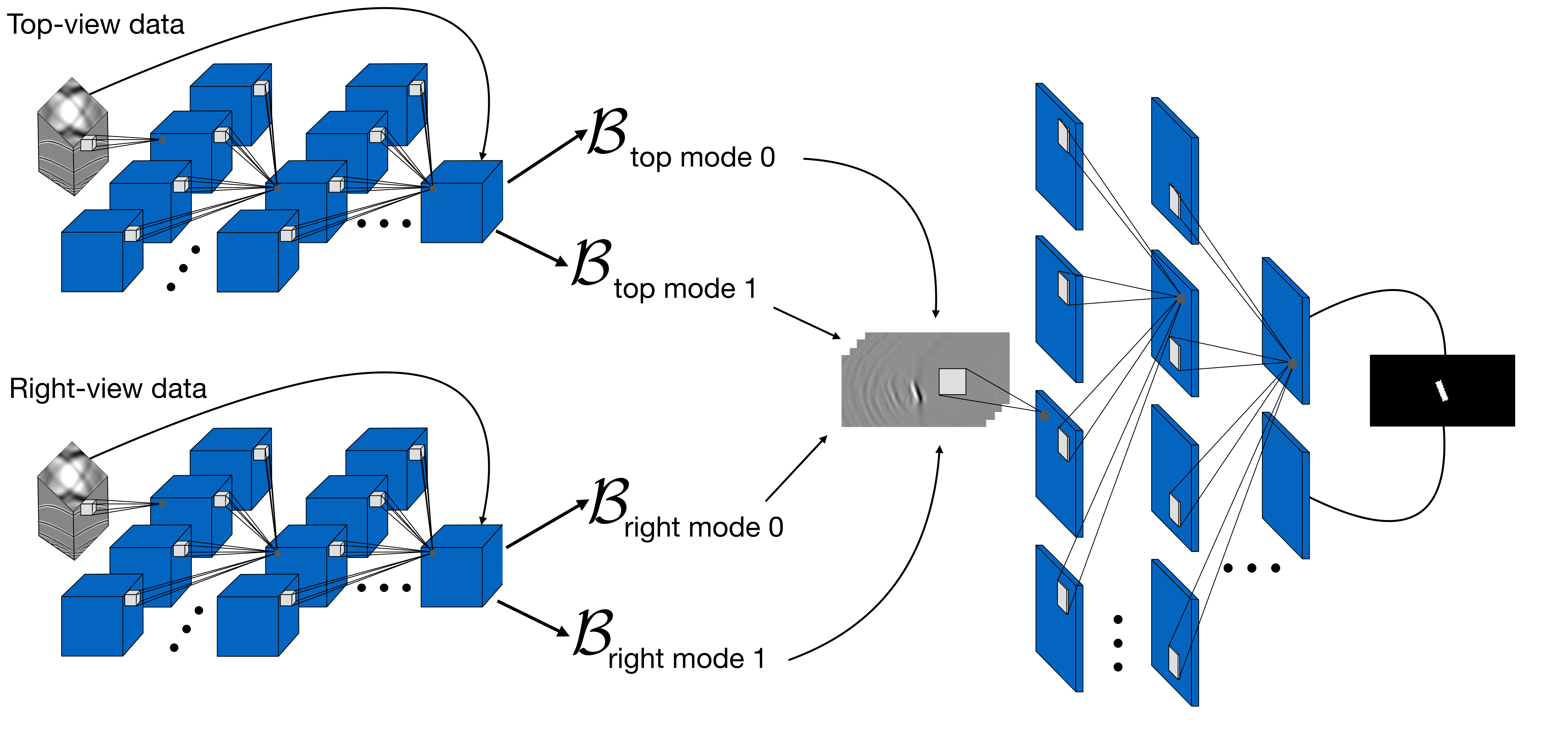}
\caption{Proposed end-to-end deep learning architecture. Two data volumes from different viewpoints are received by two 3D DCNNs for data pre-processing. Then, for each data volume, two DAS operators are used, to produce four intermediate image modes. These are used by a 2D DCNN for image post-processing to fuse them into a segmentation map of materials. One filter at one location per layer is shown, with depth and width of layers included only for illustration purposes.}
\label{2}
\end{figure*}

Here, we propose a novel Deep Convolutional Neural Network (DCNN) architecture to fuse raw data from multiple insonification paths into a segmentation map as illustrated in Figure \ref{2}. To achieve this, we incorporate an image formation method within the deep learning training where we implement a DAS image formation operator for each image mode. We demonstrate that our approach improves upon traditional image fusion methods for the characterisation of the shape of a defect in a non-destructive testing scenario. In section 2, we describe the ultrasonic data acquisition considered in this paper and the DAS image formation algorithm. Then, in section 3, we introduce our proposed deep learning fusion architecture. In section 4, we include experiments on simulated data for a rotated defect and compare our approach with a traditional image fusion technique. 

\begin{figure}[ht]
\centering
\includegraphics[scale=0.15]{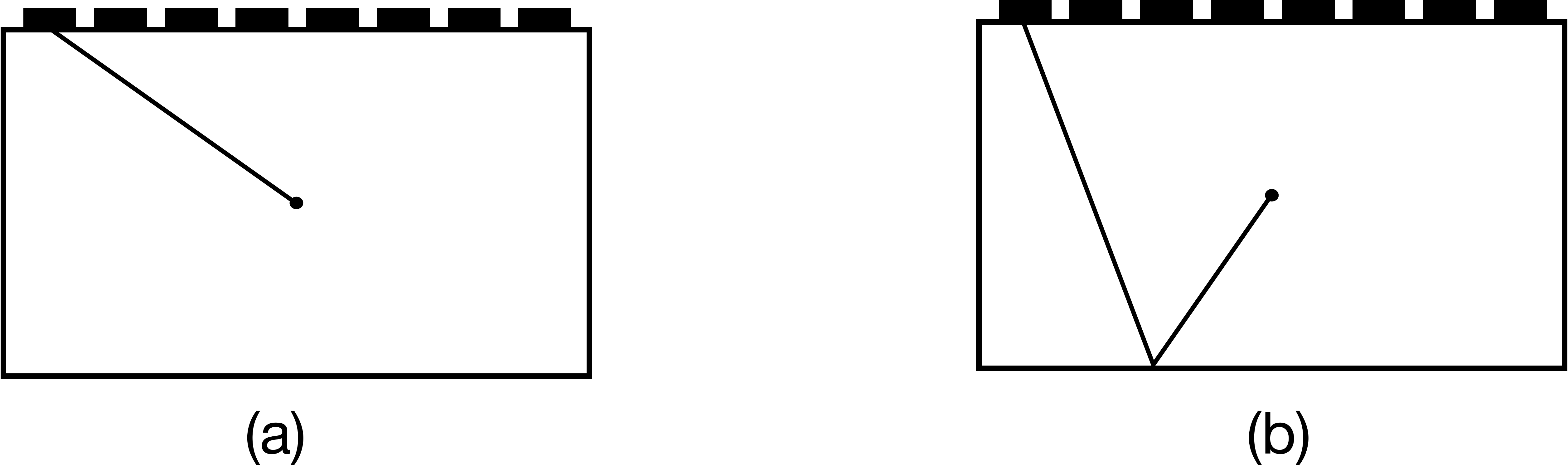}
\caption{A linear array on top of a medium with a reflective bottom boundary illustrating two different insonification paths. (a) Direct path, (b) boundary reflection path.}
\label{1}
\end{figure}
\section{Ultrasonic Data Acquisition and the Delay-And-Sum Algorithm}
In this work, we consider imaging a rectangular-shaped domain using a linear array of transducers. Data acquisition starts with a transducer acting as a source, emitting a wave into the medium and then all transducers in that linear array receive the resulting wave field. This continues with the next transducer as a source until all transducers have been used as sources \cite{8} \cite{9}. It produces a data volume, $\mathbf f \in \mathbb{R}^{n_t \times n_s \times n_r}$ where $n_t$ is the number of time samples, $n_s$ is the number of sources and $n_r$ is the number of receivers. Two reflective boundaries are used, one at the bottom and one at the left-side of the domain in order to enable the use of various insonification paths. The linear array is first placed on the top of the domain and then on its right side. This results in two data volumes, one referred to as top-view data and the other as right-view data. 

For each data volume, $\mathbf f$, the aim is to create an image, $\mathbf u \in \mathbb{R}^{n_x \times n_z}$, where $n_x$ and $n_z$ are the horizontal and vertical pixels. To achieve this, we use the Delay-And-Sum (DAS) image formation algorithm. First, we calculate the travel-times, $\tau(\mathbf r_s, \mathbf r_r, \mathbf p_i)$ between each combination of source $r_s$, receiver $r_r$ and image point $p_i$. Figure \ref{1}(a) illustrates an example of a direct path. This direct path is included in the calculation of the travel-times for an image mode which we refer to hereafter as top-mode 0. Figure \ref{1}(b) illustrates an example of a boundary reflection path. This is included in the calculation of the travel-times for image mode referred to as top-mode 1. There are many possible insonification paths but for brevity we only mention the above. For the image modes using the right-view data volumes, we refer to them as right-mode 0 and right-mode 1. Further information on how to obtain multiple image modes with various combinations of insonification paths can be found in \cite{1}. So, in total, here we consider four different image modes as an illustration, which are derived from two data volumes with different insonification paths using the corresponding travel-times. 

Once all combinations of travel-times are calculated, the time index corresponding to each travel-time is located, using the underlying sampling frequency. Then, a sum across all combinations of travel-times is performed,
\begin{equation}
u_i = \sum_{s=0}^{n_s} \sum_{r=0}^{n_r} f(\tau(\mathbf r_s, \mathbf r_r, \mathbf p_i), s, r),
\end{equation} to obtain each pixel amplitude, $u_i$, repeated for all pixels. We can compactly write the entire image formation procedure as,
\begin{equation}
\mathbf u = \mathcal{B} \mathbf f,
\end{equation} with $\mathcal{B}: \mathbb{R}^{n_t \times n_s \times n_r} \longrightarrow \mathbb{R}^{n_x \times n_z}$ referred to as the DAS operator hereafter.
\begin{figure*}
\centering
\includegraphics[scale=0.27]{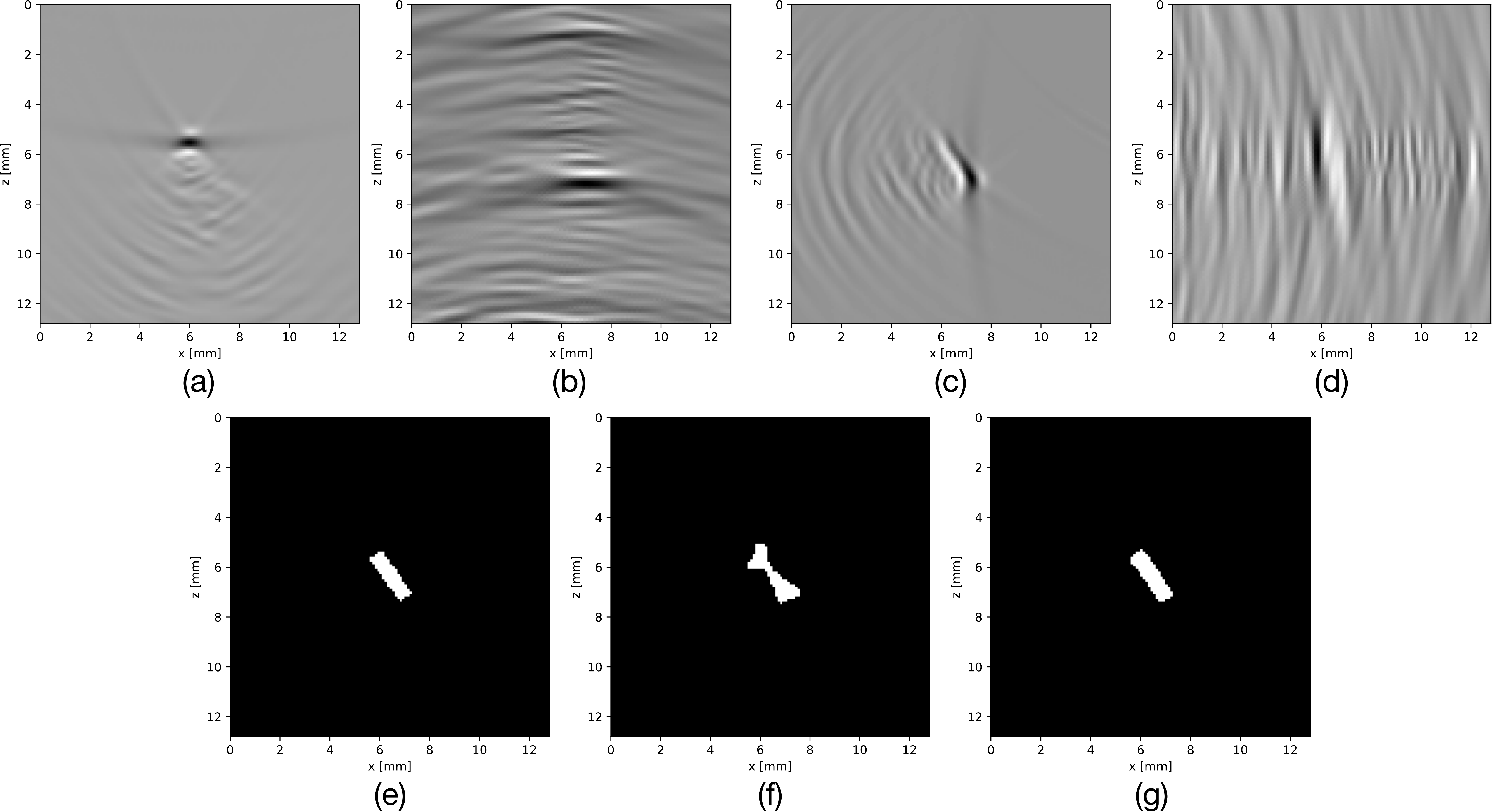}
\caption{(a) - (d): DAS image using (a) top-view data and direct path (top-mode 0), (b) top-view data and boundary reflection path (top-mode 1), (c) right-view data and direct path (right-mode 0), (d) right-view data and boundary reflection path (right-mode 1). (e) Ground truth segmentation map, (f) Traditional image fusion technique, (g) Proposed end-to-end data fusion prediction.}
\label{3}
\end{figure*}

Using the above procedure, a traditional fusion method first calculates all image modes separately. Then, it calculates the maximum amplitude per pixel across all image modes and uses this as the fused pixel per location. Edge detection algorithms or the Hilbert transform could be applied on each image mode, before calculating the maximum pixel value. In addition, certain thresholding operators could be used to discard lower amplitudes and noise. All these require manual choice of algorithmic parameters and pre-defined image transforms in order to fuse images. In this work, we propose to use end-to-end deep learning to fuse data directly into a segmentation.

\section{Proposed deep-learning fusion method}
We propose a novel Deep Convolutional Neural Network (DCNN) architecture that performs end-to-end data fusion. First, two 3D DCNNs are used for data pre-processing. Each acts on a separate data volume (top-view and right-view data) and learns to optimally process the data for the image formation step. Following each 3D DCNN, two DAS operators are used. These are implemented as a network layer that applies the DAS algorithm on the processed data. To achieve this, we also derive and code its adjoint program to enable the backpropagation of errors during training. In total, we implemented four DAS operators, each corresponding to a different insonification path. These correspond to the use of the direct or boundary reflection and for the two different viewpoints. This creates four intermediate image modes within the deep learning training, namely the top-mode 0, top-mode 1, right-mode 0 and right-mode 1. 

An image post-processing 2D DCNN receives these image modes and learns optimal filters to obtain the final segmentation. Due to the use of two data volumes in GPU memory simultaneously, one training sample per mini-batch is used. In addition, both 3D DCNNs are comprised of only 2 layers and 2 channels per layer due to memory limitations. Better performance could be achieved with a more expressive (deeper) model. The 2D DCNN is comprised of 8 layers and 16 channels per layer. Weight Standardization \cite{10} and Group Normalization \cite{11} are used per layer to improve the stability of training. Skip connections in the data pre-processing DCNNs further help information flow and decrease the required training time \cite{15}. Figure \ref{2} illustrates the proposed end-to-end data fusion architecture.

\section{Experiments}
In order to evaluate our proposed method, we examine a scenario mimicking the ultrasonic non-destructive inspection of pipelines. We simulate ultrasonic data using the k-Wave toolbox \cite{12}. The pipeline was modelled as carbon steel, with speed of sound equal to $5920$ m/s. The defect and walls are modelled as water with speed of sound equal to $1500$ m/s. Their respective densities are also taken into consideration during wave simulations. 

Two separate wave simulations are performed, that is, one per location of the linear array of transducers (top-view and right-view). This results in two data volumes, each with dimensions $2865 \times 128 \times 128$ where 128 transducers are used with pitch equal to $0.2$ mm. The time samples are equal to $2865$ with sampling frequency equal to 98.67 MHz. The domain of the medium is set to $256\times 256$, with $0.1$ mm spacing. In the middle of the domain, a defect is placed and rotated with random angles to create $220$ different scenarios. Reflective boundaries are placed on the left and bottom-side of the domain. The domain is cropped in the middle to $128\times 128$ for the medium and $1600\times 128\times 128$ for each data volume due to memory constraints. Furthermore, the materials in the segmentation maps are set to either 0 or 1, corresponding to the two different speeds of sound. Figure \ref{3}(e) includes an example of a scenario. From the 220 different scenarios, we use $200$ for training and $20$ for testing. For each scenario, two data volumes are used as input and the corresponding segmentation map is used as the target. Our proposed approach is implemented in PyTorch \cite{13} and the DCNN parameters are optimized using the Adam optimization \cite{14} with a learning rate of $10^{-3}$. The cross entropy loss function is used for training. 

For comparisons, we illustrate the four image modes produced by using the corresponding DAS operators and a result from a traditional image fusion technique. We found empirically that using the direct modes only produced better results for these scenarios using the following image fusion technique. First, we perform the Hilbert transform on the direct image modes. Then, we obtain the maximum value per pixel and use simple thresholding for segmentation. Figure \ref{3}(a) - \ref{3}(d) include the top-mode 0 (direct reflection and top-view data), top-mode 1 (boundary reflection and top-view data), right-mode 0 (direct reflection and right-view data) and right-mode 1 (boundary reflection and right-view data) respectively, obtained as discussed in section 2. Figure \ref{3}(e) includes the ground truth used for wave simulations. Figure \ref{3}(f) is the result of the traditional image fusion method and Figure \ref{3}(g) is the result of our proposed deep learning approach. 

We can see that the four image modes are able to image different parts of the defect from different angles. However, if they are used independently, it is challenging to obtain the correct location of the defect. The traditional image fusion technique is able to localize the defect but the shape is not very accurate. On the other hand, our proposed approach is able to very accurately localize and identify the correct shape of the defect. We calculated the average cross entropy (the lower the better) over 20 test samples for both our proposed approach and the traditional image fusion technique. The average cross entropy of our method is 0.003 as opposed to 0.398 for the traditional image fusion technique. This illustrates that our proposed method is visually and quantitatively better but further experiments are required to identify its limitations to more complicated and realistic scenarios. In addition, there are many possible insonification paths and combinations that could be used. Further investigation on the importance of each insonification path and on the information they provide is needed for accurate fusion of data to a segmentation map.

\section{Conclusion}
Ultrasonic data acquisition from different viewpoints offers complementary information for improved image formation and segmentation. Different insonification paths can be used corresponding to different location of transducers or different paths from reflection boundaries. This creates numerous image modes of the same underlying medium of interest that can be used for the segmentation of its materials. Traditionally, human expertise and tedious data and image processing were necessary to exploit this additional information content. In this work, we proposed an end-to-end deep learning architecture that performs data fusion directly. That is, raw ultrasonic data acquired from two different viewpoints are turned into a segmentation map of the underlying medium. This is achieved by implementing the DAS image formation algorithm into a network layer that connects data pre-processing and image post-processing DCNNs. Four different DAS operators are incorporated within the deep learning training, each for a unique insonification path. These operators constrain the deep learning training to learn convolutional filters that are optimal for the DAS image formation process. Experiments have shown that our proposed approach outperforms traditional image fusion for a non-destructive testing scenario. This illustrates the applicability of deep learning for ultrasonic data fusion.

\bibliographystyle{IEEEtran}
\bibliography{references}

\end{document}